\documentclass[12pt]{article}
  \usepackage[dvips]{graphicx}
\usepackage{amsmath}
\newcommand{\nc}{\newcommand}
\nc{\postscript}[2] 
{\setlength{\epsfxsize}{#2\hsize}\centerline{\epsfbox{#1}}}
\nc{\non}{\nonumber}
\nc{\pt}{p_{{}_T}}
\nc{\lmc}{\Lambda_c^+}
\nc{\plmc}{\vec{\Lambda}_c^+}
\nc{\all}{D_{LL}}
\nc{\dg}{\Delta G(x,Q^2)}
\nc{\stil}{\tilde{s}}
\nc{\ttil}{\tilde{t}}
\nc{\util}{\tilde{u}}
\nc{\shs}{\hat{s}}
\nc{\ths}{\hat{t}_1}
\nc{\uhs}{\hat{u}_1}
\nc{\cosec}{\rm cosec}
\nc{\mpr}{m_p}
\nc{\mc}{m_c}
\nc{\mlc}{m_{\lmc}}
\nc{\pa}{p_{{}_a}}
\nc{\pb}{p_{{}_b}}
\nc{\pA}{p_{{}_A}}
\nc{\pB}{p_{{}_B}}
\nc{\plc}{p_{\lmc}}
\nc{\pc}{p_{{}_c}}
\nc{\dsp}{D^{*+}}
\nc{\dspm}{D^{*\pm}}
\nc{\bm}[1]{\mbox{\boldmath $#1$}}

\def\dps{\displaystyle}
\def\mib#1{\mbox{\boldmath $#1$}}

\nc{\prd}[3]{{\it Phys.\ Rev.}\ {{\bf D{#1}} (#2), #3}}
\nc{\prl}[3]{{\it Phys.\ Rev.\ Lett.}\ {{\bf {#1}} (#2), #3}}
\nc{\plb}[3]{{\it Phys.\ Lett.}\ {{\bf B{#1}} (#2), #3}}
\nc{\npb}[3]{{\it Nucl.\ Phys.}\ {{\bf B{#1}} (#2), #3}}
\nc{\ptp}[3]{{\it Prog.\ Theor.\ Phys.}\ {{\bf {#1}} (#2), #3}}
\nc{\zfp}[3]{{\it Z.\ Phys.}\ {{\bf C{#1}} (#2), #3}}
\nc{\mpla}[3]{{\it Mod.\ Phys.\ Lett.}\ {{\bf A{#1}} (#2), #3}}
\nc{\rmp}[3]{{\it Rev.\ Mod.\ Phys.}\ {{\bf {#1}} (#2), #3}}
\nc{\ijmpa}[3]{{\it Int.\ J.\ of\ Mod.\ Phys.}\
               {{\bf A{#1}} (#2), #3}}
\nc{\epj}[3]{{\it Eur.\ Phys.\ J.}\ {{\bf C{#1}}  (#2), #3}}
\begin{document}
\pagestyle{empty} \setlength{\footskip}{2.0cm}
\setlength{\oddsidemargin}{0.5cm} \setlength{\evensidemargin}{0.5cm}
\renewcommand{\thepage}{-- \arabic{page} --}
\def\mib#1{\mbox{\boldmath $#1$}}
\def\bra#1{\langle #1 |}      \def\ket#1{|#1\rangle}
\def\vev#1{\langle #1\rangle} \def\dps{\displaystyle}
%
   \def\thebibliography#1{\centerline{\Large \bf REFERENCES}
     \list{[\arabic{enumi}]}{\settowidth\labelwidth{[#1]}\leftmargin
     \labelwidth\advance\leftmargin\labelsep\usecounter{enumi}}
     \def\newblock{\hskip .11em plus .33em minus -.07em}\sloppy
     \clubpenalty4000\widowpenalty4000\sfcode`\.=1000\relax}\let
     \endthebibliography=\endlist
   \def\sec#1{\addtocounter{section}{1}\section*{\hspace*{-0.72cm}
     \normalsize\bf\arabic{section}.$\;$#1}\vspace*{-0.3cm}}
\renewcommand{\thesection}{\arabic{section}.$\!\!$}
\vspace*{-1.8cm}
\begin{flushright}
$\vcenter{
\hbox{KOBE-FHD-03-02}
\hbox{FUT-03-01}
\hbox{hep-ph/0306285}
}$
\end{flushright}
\renewcommand{\thefootnote}{$\dag$}
\vskip 0.2cm
\begin{center}
{\Large \bf Probing Polarized Gluon Distributions through
Charmed Hadron Production in Polarized \bm{pp} Collisions
at BNL--RHIC Energy \\
\vskip 0.5cm
}

\vskip 0.15cm
\end{center}

\vspace*{0.2cm}
\begin{center}
\renewcommand{\thefootnote}{\alph{footnote})}
 \setcounter{footnote}{0}
{\sc \large Toshiyuki MORII}\vspace*{0.2cm}\\
Faculty of Human Development,\\
Kobe University, \\
Nada, Kobe 657-8501, JAPAN\\
Electronic address{\tt :morii@kobe-u.ac.jp}\vspace{1.0cm}\\

{\sc \large Kazumasa OHKUMA}\vspace*{0.2cm}\\
Department of Information Science,\\
Fukui University of Technology,\\
 Gakuen, Fukui, 910-8505, JAPAN\\
Electronic address{\tt :ohkuma@fukui-ut.ac.jp}
\end{center}
\vspace*{0.8cm}
\centerline{\large ABSTRACT}

\vspace*{0.2cm}
\baselineskip=15pt plus 0.1pt minus 0.1pt
{
In order to extract information about behavior of 
polarized gluons in the nucleon, charmed hadron productions, 
i.e. $\dsp$ meson and $\lmc$ baryon productions, are studied 
in polarized $pp$ reactions at BNL-RHIC energy.
For these processes, the spin correlation asymmetry $D_{LL}$ between
the target proton and the produced charmed hadron,
and its statistical sensitivity $\delta D_{LL}$ are calculated.
From analyses on these processes, 
we found that the  pseudo-rapidity  distribution of $D_{LL}$ 
in the limited transverse momentum  region 
is quite effective for distinguishing the model of polarized 
gluons as well as the model of spin-dependent fragmentation functions.\\
\vfill
}
{\footnotesize
\noindent
{\it PACS number(s)}: 
13.85.Ni. 
13.88.+e, 
14.20.Lq, 
14.40.Lb, 

\vspace*{-0.2cm}\\
\phantom{Keywords:~}Charmed Hadron Production.
}
\newpage

\renewcommand{\thefootnote}{$\sharp$\arabic{footnote}}
\pagestyle{plain} \setcounter{footnote}{0}
\pagestyle{plain} \setcounter{page}{1}
\baselineskip=18.0pt plus 0.2pt minus 0.1pt
\section{Introduction}	

~~Though the proton is one of the most familiar nucleons,
the origin of its spin is still an open question.
About 15 years ago, the European Muon Collaboration (EMC) reported 
surprising data on the polarized structure function of proton
$g^p_1(x,Q^2)$~\cite{emc}, indicating that the spin of 
the proton cannot be described by the naive quark model.
Disagreement of the result extracted from the EMC data with 
theoretical predictions has been called the 
``{\it Proton Spin Puzzle}'', which is not yet solved well 
in spite of many follow-up experiments and a great development 
of theoretical analyses.  It is still one of the 
challenging themes in nuclear and particle physics~\cite{rpsp}. 
According to the parton-model concept inspired by quantum 
chromodynamics(QCD), the proton spin satisfies the following
sum rule; 
\begin{equation}
\frac{1}{2}=\int_0^1 dx
\left[
\frac{1}{2}\sum_q (\Delta q (x)+ \Delta \bar{q}(x)) + \Delta g(x) 
\right]
+<L>_{q+g}
\label{spin_sum}
\end{equation}
where $\frac{1}{2}$ on the left-hand side is a spin of proton,
while $\Delta q(x), \Delta \bar{q}(x)$ and $\Delta g(x)$ are
the spin carried by quarks, anti-quarks and gluons, respectively.
In the right-hand side of Eq.(\ref{spin_sum}), integration is
performed over all Bjorken-$x$ region.  
$<L>_{q+g}$ represents the orbital angular momentum of quarks and 
gluons in the proton.  An important thing is to know how large each 
of these component is and then to understand, based on QCD, what the 
underlying dynamics of this sum rule is. 
These days the behavior of valence quarks in the proton 
has been considerably well understood with great efforts.
However, the knowledge of gluons in the proton is still poor. 
Therefore, to solve the proton spin puzzle, it is very important 
to obtain a good information about gluon polarization 
in the proton.
So far, the gluon distribution in the proton 
has been studied mainly through deep inelastic scattering (DIS) 
of polarized leptons (electrons and muons) off polarized nucleons.
However, now we are in a new stage; the Relativistic Heavy 
Ion Collider (RHIC) at Brookhaven National Laboratory (BNL) which gives 
us another unique place for probing the proton structure via 
polarized proton--polarized proton scattering has started to work.
For the RHIC experiment, the following  
processes are already proposed to study the behavior of polarized gluons 
in the proton~\cite{rhic};
 \begin{itemize}
  \item High-$\pt$ (``prompt'') photon production, 
        $\vec{p}\vec{p}\to\gamma X$, 
  \item Jet production, 
        $\vec{p}\vec{p}\to jet(s) X$,
  \item Heavy-flavor production, 
         $\vec{p}\vec{p}\to c\bar{c} X,b\bar{b}X$,
 \end{itemize}
where the particle with an arrow means that it is polarized.

\begin{figure}[htb]
\includegraphics[scale=0.46]{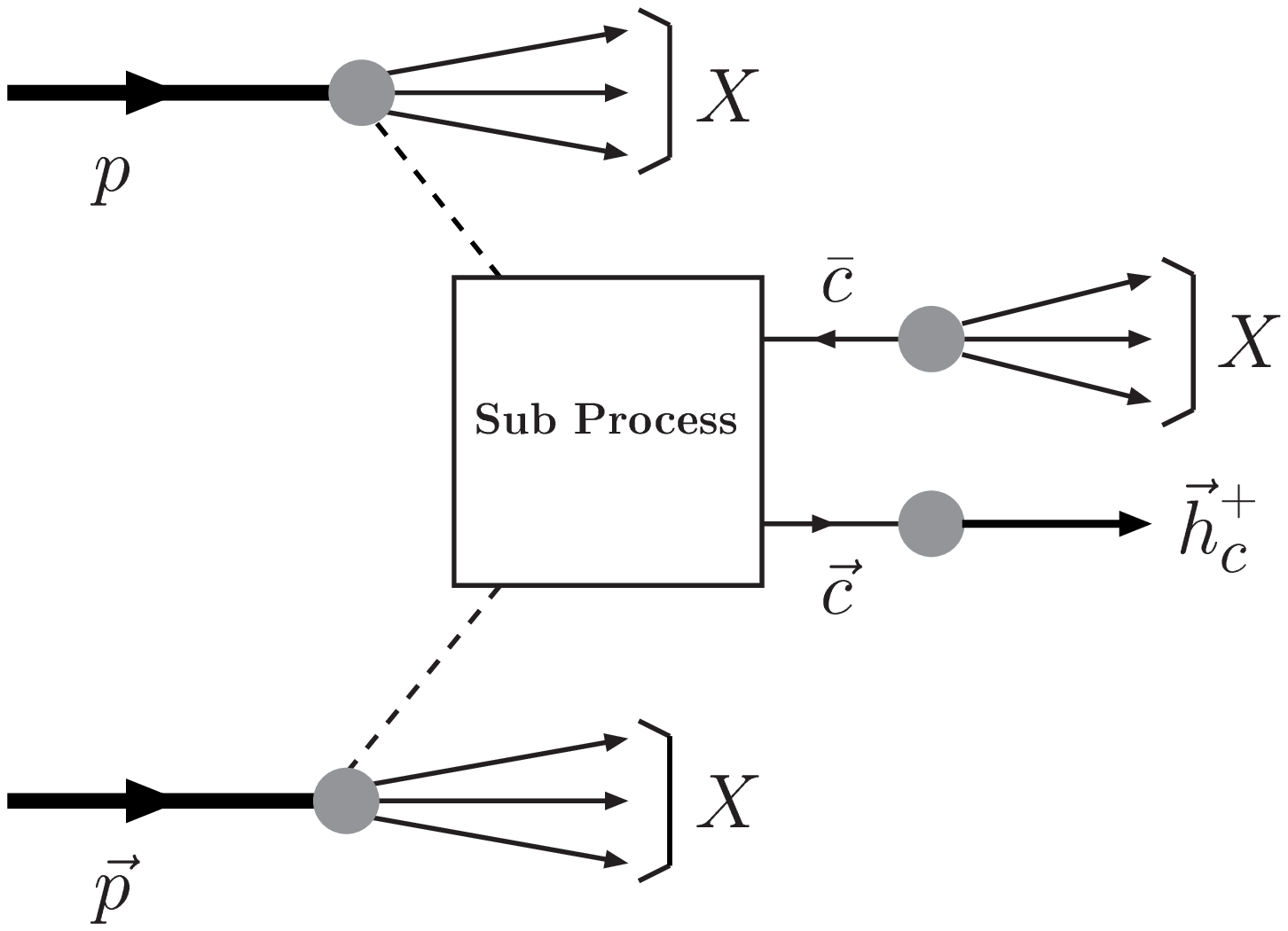}
\hspace*{0.2cm}
\includegraphics[scale=0.5]{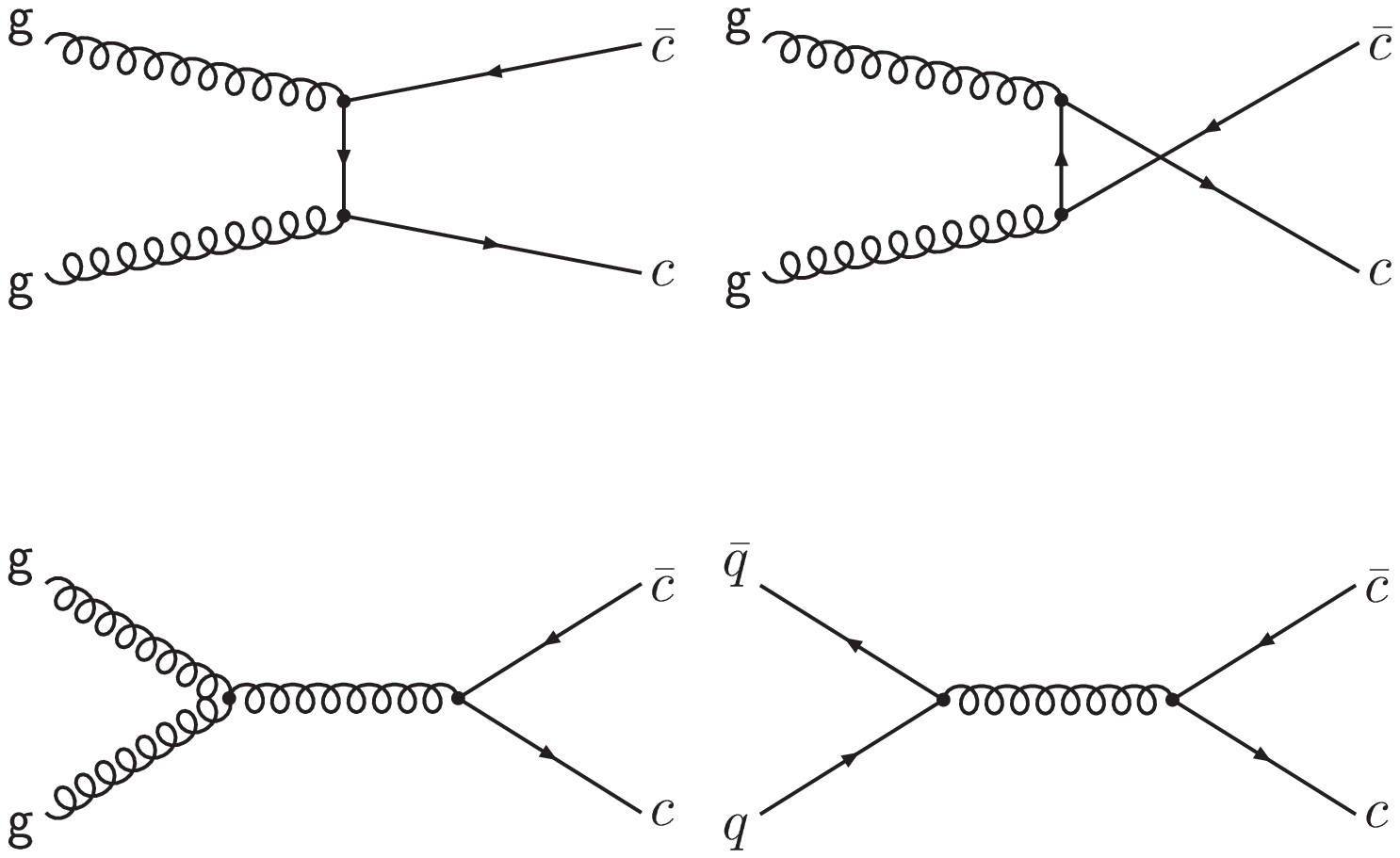}
 \caption{
Diagrams for $ p\vec{p}\to\vec{h}_c^{{}^{}_+}X$ 
at the lowest order (left) and its subprocesses (right), 
where $\vec{h}_c^+$ means a polarized charmed hadron with positive
charge. Dashed-lines in the left figure stand for
partons which are emitted from initial protons.  }
\label{main}
\end{figure}

In addition to these processes,
we propose other interesting processes, i.e. the polarized charmed 
hadron productions in the polarized proton--unpolarized proton collision: 
\begin{eqnarray}
&&p\vec{p}\to c\bar{c}\to\vec{D}^{*+}X,\\
&&p\vec{p}\to c\bar{c}\to\vec{\Lambda}_c^{+}X,
\end{eqnarray}
by expecting that they also could be observed in
the forthcoming RHIC experiment.  
In the lowest order of QCD, a charm quark being one of the constituents
of charmed hadrons is produced via gluon--gluon fusion 
($gg \to c\bar{c}$) and quark--anti-quark annihilation 
($q\bar{q} \to c\bar{c}$) processes
in proton--proton collisions (Fig.\ref{main}).
Then, if we could separate kinematically a contribution of 
$gg \to c\bar{c}$ from $q\bar{q} \to c\bar{c}$ and pick up
$gg \to c\bar{c}$ clearly, 
we could extract a good information about gluon 
polarization by observing the spin correlation asymmetries 
between the initial proton and these produced charmed hadrons,
since the cross section of gluon fusion
is directly proportional to the gluon distribution. 

Concerning the $\Lambda_c^+$ production, in a couple of years ago
we analyzed the spin correlation between the initial proton
and produced $\Lambda_c^+$  in the diffractive regions of the process
$p\vec{p}\to p \vec{\Lambda}^+_c X$, where Pomeron interactions 
play an important role~\cite{om} and pointed out that 
the process is promising for extracting the polarized gluon distribution.
In this paper, we develop those analyses to another important 
kinematical region, i.e. the central region of collision.  The present
analysis is complementary to those previous analyses and should give us 
additional information on the polarized gluon distribution in the 
different kinematical region~\footnote{
The preliminary analysis of the $\lmc$ production was done in
ref~\cite{osm}.
}.

This paper is organized as follows.
In the next section, we explain why we focus on charmed hadron 
production in this work. 
Then, we introduce spin correlation observable and present
a theoretical formulation of our calculation in Sec.~3. 
In section~4, we show the result of our numerical calculation.
Finally, Sec.~5 is devoted to conclusion, including discussion
and summary.

\section{Polarized charmed hadron production in \bm{pp} collisions}

~~As described above, we will focus on two processes of charmed hadron
production; one is polarized $\dsp$ meson production and another
is polarized $\lmc$ production. 
Then, what is so special about these processes? and why are they  
so interesting?  This is because in these processes, we can regard 
the spin of the produced charmed hadron to be same as 
the one of the charm quark as described below; 
\begin{itemize}
\item $\dsp$ meson\\
Since the $\dsp$ meson is in the ${}^3S_1$ state in the non-relativistic 
quark model, the spin of $\dsp$ meson is carried by a charm 
quark and an anti-down quark whose spins are combined in parallel.
Therefore, the spin direction of the $\dsp$ meson and
the spin direction of the constituent charm quark  
are expected to be same.

\item $\lmc$ baryon\\
The $\lmc$ is composed of a heavy charm quark  and antisymmetrically
combined light up  and down quarks.  
Hence, the $\lmc$ spin is basically carried by a charm quark.
\end{itemize}
Furthermore, it is expected to be very rare for a produced charm quark 
to change its spin direction during its decay into a charmed hadron,
since the charm quark is significantly heavy and the spin flip interaction
being proportional to $1/m_c$ ($m_c$ is charm quark mass) is small.
Therefore, if these processes are originated from gluon fusion,
spin-dependent observables for produced charmed hadrons directly 
depend on the polarization of gluons in the initial proton and thus,  
observation of the polarization of the produced
charmed hadron could give us a good information about 
polarized gluons in the proton.

Concerning the above discussion, it is worth while to comment on
contribution of light-quark fragmentation to charmed hadron 
production.
In the case of charmed hadron production by light-quark fragmentation, 
a light-quark produced in a hard parton subprocess must pick up
a charm quark created from vacuum to make a charmed hadron.
However, the probability of charm-quark pair creation is extremely 
small, estimated to be $10^{-11}$ compared to the one for
 $u\bar u$ and $d\bar d$
pair creation~\cite{Lund}.
Therefore, in charmed hadron production we can safely neglect the contribution
of light-quark fragmentation.  This is rather different from the case of
$\Lambda$ production~\cite{ma}, where the probability of $s\bar s$ pair creation from vacuum
cannot be neglected.  
\section{Spin correlation observables and cross sections}
To study the polarized gluon distribution in the proton, we introduce
the spin correlation asymmetry between the polarized target-proton and
the produced charmed hadron defined by;
\begin{eqnarray}
D_{LL}&=&\frac{d \sigma_{++} - d\sigma_{+-} + d \sigma_{--}- d\sigma_{-+}}
{d \sigma_{++} + d \sigma_{+-}+d \sigma_{--} + d \sigma_{-+}}\nonumber\\
&\equiv&\frac{{d \Delta \sigma}/{d X}}
{{{d\sigma}/{d X}}},~~(X=\pt~{\rm or}~\eta), \label{all}
\end{eqnarray}
where $d \sigma_{+ -}$, for example, denotes the spin-dependent
differential cross section with the positive helicity of the target proton
and the negative helicity of the produced charmed hadron.
$\pt$ and $\eta$, which are represented by $X$ in Eq.(\ref{all}),
are transverse momentum and pseudo-rapidity of
the produced charmed hadron, respectively. 

In order to perform our analyses, we take the proton--proton 
center of mass system(CMS). 
In this system, four momentum $p_{{}_i}$ of each particle $i$
is defined as follows;
\begin{eqnarray}
&&p_{{}_{A/B}}=\frac{\sqrt{s}}{2}
\left(1,~\mp \beta,~\bm{0}
\right)
~,~~~~~\beta 
\equiv \sqrt{1-\frac{4 m^2_{{}_p}}{s}} \non \\
&&p_{{}_{h_c}}\equiv(E_{{}_{h_c}},~p_{{}_L},~\bm{p_{{}_T}}) 
\label{cmsv}\\
&&\phantom{p_{{}_{h_c}}}=(\sqrt{m_{h_c}^2+{p_{{}_{T}}}^2
{\rm cosh^2}\eta},~
{p_{{}_T}}{\rm sinh}\eta,~\bm{p_{{}_T}}),\non \\
&&p_{{}_{a/b}}=x_{{}_{a/b}}p_{{}_{A/B}},~~~p_{c}=\frac{p_{{}_{h_c}}}{z}, \non
\end{eqnarray}
where the parameters in parentheses denote 
the energy, the longitudinal momentum and
the transverse momentum, in this order. 
$m_i$ stands for the mass of the particle $i$ 
($i=p$ for proton and $i=h_c$ for $\lmc~{\rm or}~ \dsp$).
$x_{{a/b}}$ and $z$ are momentum fractions of the parton to  
the proton and 
of the charmed hadron to the charm quark, respectively.  
Here, notice that we define the momentum of unpolarized and polarized proton
as $p_{{}_A}$ and $p_{{}_B}$, respectively.
In addition, we regard the direction of $p_{{}_B}$ as  
the positive $z$--axis. 

According to the quark-parton model,
$p_{{}_T}$- or $\eta$-dependent differential
cross section, $d \Delta \sigma / d X$, is given by,
\footnote{
In Eq.(\ref{eq:dcross}), $X$ or $Y$ mean $p_{{}_T}$ or $\eta$.
Thus, if $X$ is $p_{{}_T}$, then $Y$ is $\eta$, and vice versa.}
\begin{eqnarray}
&&\frac{d \Delta \sigma}{dX}
 = \int^{Y{{}^{\rm max}}}_{Y{{}^{\rm min}}}
\int^{1}_{x^{{}^{\rm min}}_{{}_b}} 
\int^{1}_{x^{{}^{\rm min}}_{{}_a}}
\left[
   g_{g_{{}_a}/p_{{}_A}}(x_a,Q^2)
   \Delta g_{\vec{g}_{{}_b}/\vec{p}_{{}_B}}(x_b,Q^2) 
\frac{d \Delta \hat{\sigma} (gg\to c\bar{c})}{d \hat{t}}
\right. \non \\
&&
\phantom{\frac{d{\sigma}}{d\pt}
}
+\sum_{q=u,d,s,\bar{u},\bar{d},\bar{s}}
\left\{
q_{q_{{}_a}/p_{{}_A}}(x_a,Q^2)
\Delta q_{\vec{q}_{{}_b}/\vec{p}_{{}_B}}(x_b,Q^2)\right.
\non \\
&&
\phantom{d\frac{\hat{\sigma}}{dX}
\sum_{q=u,d,s,\bar{u}}
}
\left. +\Delta q_{q_{{}_a}/p_{{}_A}}(x_a,Q^2)
q_{\vec{q}_{{}_b}/\vec{p}_{{}_B}}(x_b,Q^2)
\right\}
\left.
\frac{d \Delta \hat{\sigma}(q\bar{q}\to c\bar{c})}{d \hat{t}}
\right]\label{eq:dcross}\\
&&
\phantom{d\frac{\hat{\sigma}}{dX}\int_a^b\int_a^b\sigma\sigma}
\times
\Delta {D}_{\vec{h}_c^+/\vec{c}}(z)J
dx_a dx_b dY, 
~~~\left(X,Y = \pt~{\rm or}~\eta ~(X \neq Y)\right)
\non
\end{eqnarray}
where 
$ g_{g_a/p_{{}_A}}(x_a,Q^2)$ and 
$\Delta g_{\vec{g}_{{}_b}/\vec{p}_{{}_B}}(x_b,Q^2)$ are the 
unpolarized and polarized gluon distribution functions, respectively, 
and $q_{q_a(\vec q_b)/p_A(\vec p_B)}(x_{a (b)}, Q^2)$ 
and
$\Delta q_{q_a(\vec q_b)/p_A(\vec p_B)}(x_{a (b)}, Q^2)$ 
denote the unpolarized and polarized distributions, respectively, 
of the quark and anti-quark.
$\Delta {D}_{\vec{h}_c^+/\vec{c}}(z)$  
represents the spin-dependent fragmentation function of the
outgoing charm quark decaying into a polarized charmed hadron.
The spin-dependent differential cross sections of the subprocess
is given by
\footnote{
These algebraic calculations are carried out using FORM~\cite{form}.}
\begin{eqnarray}
\frac{d \Delta\hat{\sigma}(gg \to c\bar{c})}{d \hat{t}}
=\frac{\pi \alpha^2_s}{\hat{s}^2}
\left[ \frac{m_c^2}{24} \left\{
\frac{9 \hat{t}_1 -19 \hat{u}_1}{\hat{t}_1 \hat{u}_1}
+\frac{8 \hat{s}}{\hat{u}_1^2}\right\}
+\frac{\hat{s}}{6}
\left\{
\frac{\hat{t}_1-\hat{u}_1}{\hat{t}_1 \hat{u}_1}
\right\}
-\frac{3}{8}\left\{ \frac{2 \hat{t}}{\hat{s}} +1 \right\}
\right],
\label{cross}
\end{eqnarray}
\begin{equation}
\frac{d \Delta\hat{\sigma}(q\bar{q} \to c\bar{c})}{d \hat{t}}
=\frac{\pi \alpha^2_s}{\hat{s}^4}\frac{4}{9}
\left\{2m_c^2 (\hat{u}_1-\hat{t}_1)-(\hat{u}_1^2-\hat{t}_1^2)
\right\},
\end{equation}
and the Jacobian, $J$, which transforms the variables  $z$ and $\hat{t}$ 
into $\pt$ and $\eta$, is given by
$$
J\equiv \frac{2 s \beta {p}_{{}_T}^2{\rm cosh}\eta}{z \hat{s}\sqrt{m_c^2+
\pt^2{\rm cosh^2 \eta}}},
$$
where we defined the following variables;
\begin{equation}
\hat{s}=x_a x_b s,\ \ \
\hat{t}_1\equiv \frac{x_b}{z}\tilde{t},\ \ \ 
\hat{u}_1\equiv \frac{x_a}{z}\tilde{u}, \label{mandel2}
\end{equation}
with
\begin{eqnarray*}
&&s\equiv(p_{{}_A}+p_{{}_B})^2,\\
&&\tilde{t} \equiv (p_{{}_B}-p_{{}_{h_c}})^2-m_p^2-m_{{}_{h_c}}^2
=-\sqrt{s}\left[
\sqrt{m_{{}_{hc}}^2+p_{{}_T}^2 {\rm cosh}^2\eta }-\beta p_{{}_T} 
{\sinh}\eta 
\right],\non \\
&&\tilde{u} \equiv (p_{{}_A}-p_{{}_{h_c}})^2 -m_p^2-m_{{}_{h_c}}^2
=-\sqrt{s}\left[
\sqrt{m_{{}_{hc}}^2+p_{{}_T}^2 {\rm cosh}^2\eta }+\beta p_{{}_T} 
{\sinh}\eta 
\right].\non \\
\label{mandel}
\end{eqnarray*}

In Eq.(\ref{eq:dcross}), the minima of $x_a$ and $x_b$ are
given by
\begin{equation}
x^{min}_a=\frac{x_1}{1-x_2},~~~~~x^{min}_b=\frac{x_ax_2}{x_a-x_1}
\end{equation}
with
$$
 x_1 \equiv -\frac{\tilde{t}}{s-2m_p^2},~~~~~
 x_2 \equiv -\frac{\tilde{u}}{s-2m_p^2}.
$$

The unpolarized differential cross section for this process  
is calculated by replacing spin-dependent functions $\Delta g(x)$, 
$\Delta q(x)$, $\Delta D(z)$ and $\frac{d\Delta\hat{\sigma}}{dt}$ by 
spin-independent ones $g(x)$, $q(x)$, $D(z)$ and $\frac{d\hat\sigma}{dt}$,
respectively~\footnote{
The spin-independent subprocess differential cross section
$\frac{d\hat\sigma}{dt}$ was given in Ref. ~\cite{unpolamp}} 
and we use it to estimate the  denominator of $\all$.

\section{Numerical calculation}

~~To carry out a numerical calculation of $D_{LL}$, 
we used, as input parameters,
$m_c = 1.20$ GeV, $m_p = 0.938$ GeV,
$m_{{}_{\dsp}}$=2.01GeV and $\mlc = 2.28$ GeV~\cite{pdg}.
In addition, we used the AAC~\cite{aac} and  GRSV01~\cite{grsv}
parameterization models for the polarized parton distribution 
function and the GRV98~\cite{grv} model for the unpolarized one,
and set the scaling variable $Q^2$ as $Q^2=p_{{}_{T}}^2$.
Though both AAC and GRSV01 models excellently reproduce the experimental 
data on the polarized structure function of nucleons $g_1(x)$,
the polarized gluon distributions for those models are quite different. 
In other words, the data on polarized structure function of nucleons 
$g_1(x)$ alone are not enough to distinguish the model of gluon 
distributions.
Since the process is semi-inclusive, it is necessary to know the 
fragmentation function of a charm quark to $\dsp$ or $\lmc$ to 
carry out numerical calculations.
For the spin--independent fragmentation function $D_{h_c/c}(z)$, 
we use the {\it Peterson fragmentation function} which was proposed by 
Peterson et al.~\cite{peter} a long time ago and has been widely used 
for phenomenological analyses.  
This fragmentation function include one free parameter $\epsilon_h$ which
is determined by experimental analysis.
In this work, we take $\epsilon_{D^*}$=0.078~\cite{FD}
and $\epsilon_{\Lambda_c}$=0.25~\cite{FLamc} for $\dsp$ production and
$\lmc$ production, respectively. 
Furthermore, for the spin-dependent 
fragmentation function of a charm quark decaying into polarized 
charmed hadrons, $\Delta D_{\vec{h}_c/\vec{c}}(z)$, we have no information
about it at present, since we have no data on polarized charmed 
hadron productions. 
Therefore, here we simply assume the spin--dependent fragmentation
function to be parametrized~\cite{polfrag} as
\begin{equation}
\Delta D_{\vec{h}_c/\vec{c}}(z)= C_{h_c/c} 
D_{h_c/c}(z),
\end{equation}
where $C_{h_c/c}$ is a scale-independent spin-transfer coefficient.
We consider the following two models:\\
\phantom{In }
(A) $C_{h_c/c}=1$ (non-relativistic quark model) \\
\phantom{In }
(B) $C_{h_c/c}=z$ (Jet fragmentation model~\cite{jet}).\\
According to the consideration mentioned in sec.2,  if the 
spin of the charmed hadron 
is same as the spin of the charm quark produced in
the subprocess, the model (A) might be a reasonable scenario
\footnote{
From HERMES experiment, 
even the case of $\Lambda$ baryon which include lighter
$s$ quark, a fragmentation function
based on the naive-quark-parton model seems 
to be reasonable scenario~\cite{flam}.
Therefore, the model (A) must not be also an unreasonable 
scenario for $\lmc/\dsp$ productions. 
}.
Concerning the model (B), we follow the analysis for $\Lambda$ 
production~\cite{jet} and apply it to $\lmc/\dsp$ productions.

\begin{figure}[htb]
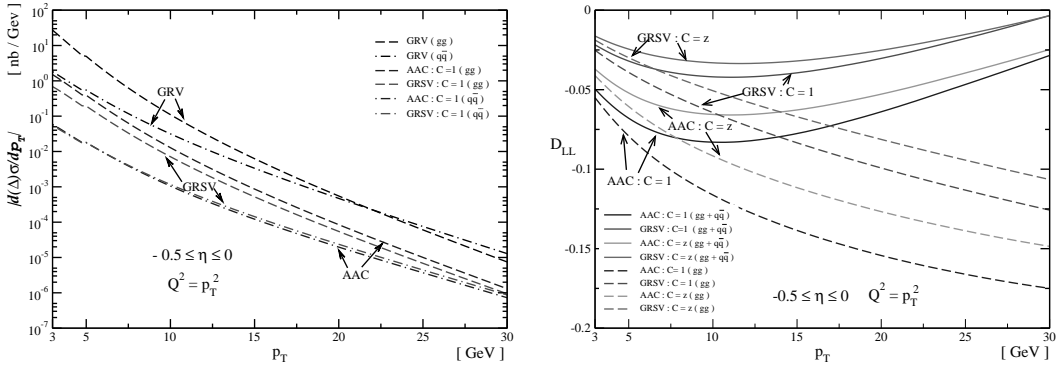

\vspace*{0.5cm}
\includegraphics[scale=0.28]{200cross_d.eps}
\hspace*{0.2cm}
\includegraphics[scale=0.28]{200dll_d_pt.eps}
  \caption{The cross sections (left panel) 
and $\all$ (right panel)
for $\dsp$ productions 
as a function of $\pt$ at $\sqrt{s}$=200~GeV.
Here, $gg$, $q\bar{q}$ and $gg+q\bar{q}$ 
shown in the parenthesis at explanatory notes
correspond to the subprocess 
taken into consideration.
}
\label{d_cross_200_pt}
\end{figure}

\begin{figure}[htb]
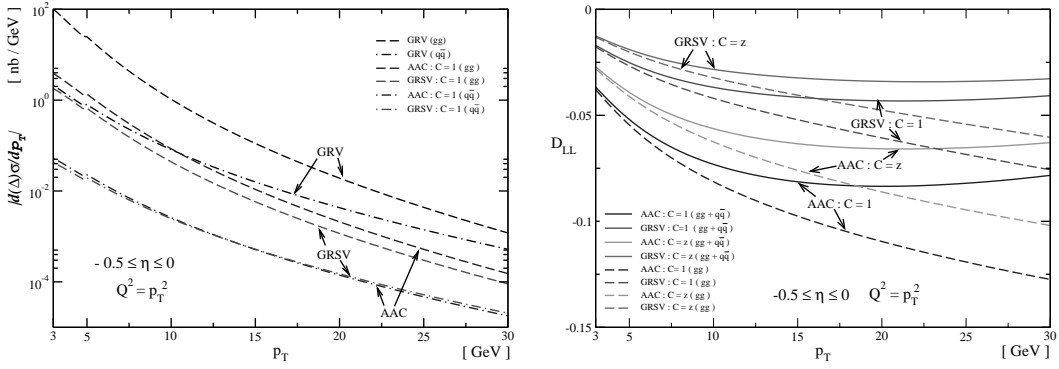

\vspace*{0.5cm}
\includegraphics[scale=0.28]{500cross_d.eps}
\hspace*{0.2cm}
\includegraphics[scale=0.28]{500dll_d_pt.eps}
  \caption{
The same as in Fig.~\ref{d_cross_200_pt},
but for $\sqrt{s}$=500GeV.
}
\label{d_cross_500_pt}
\end{figure}

\begin{figure}[htb]
\vspace*{0.5cm}
\includegraphics[scale=0.28]{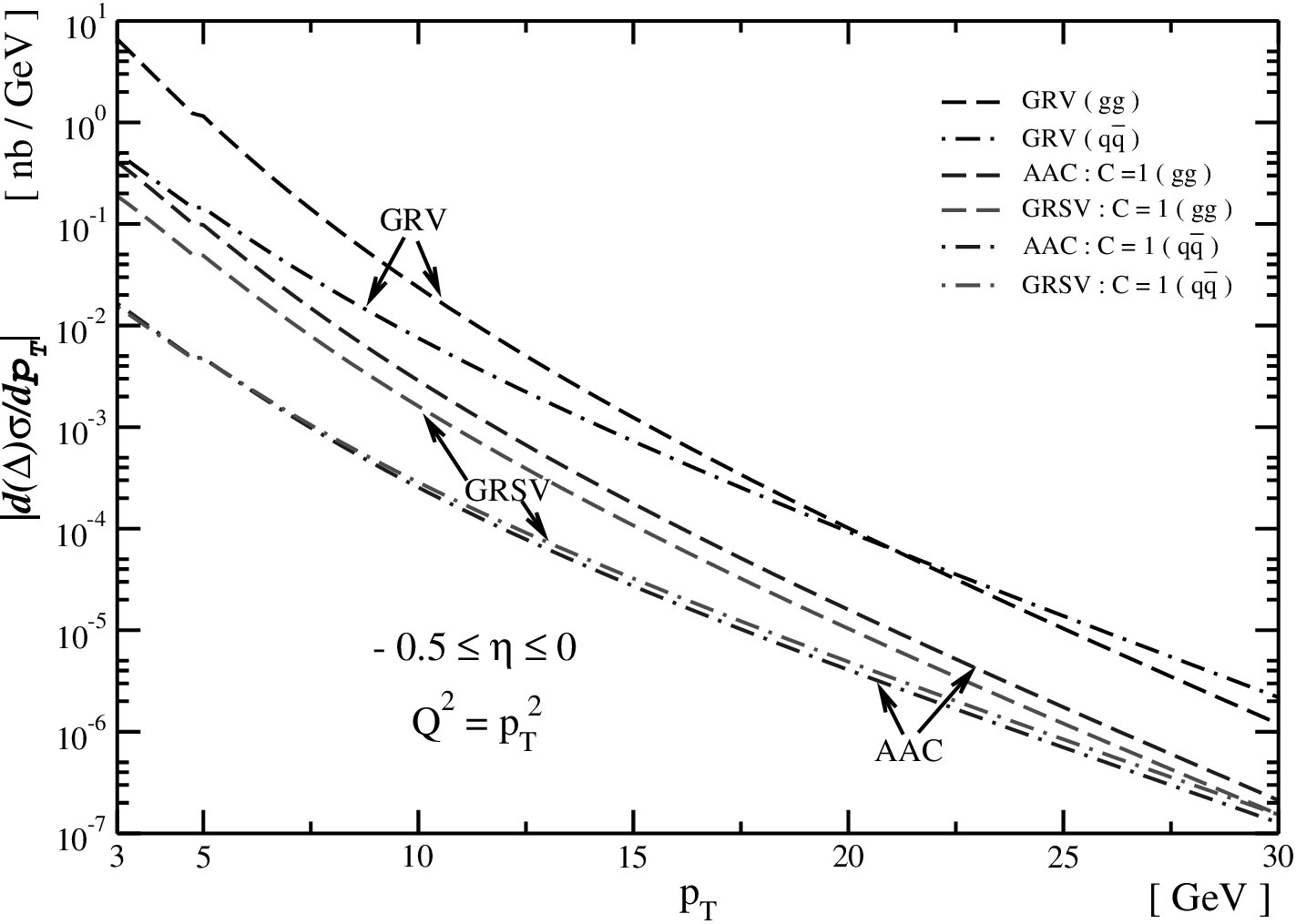}
\hspace*{0.2cm}
\includegraphics[scale=0.28]{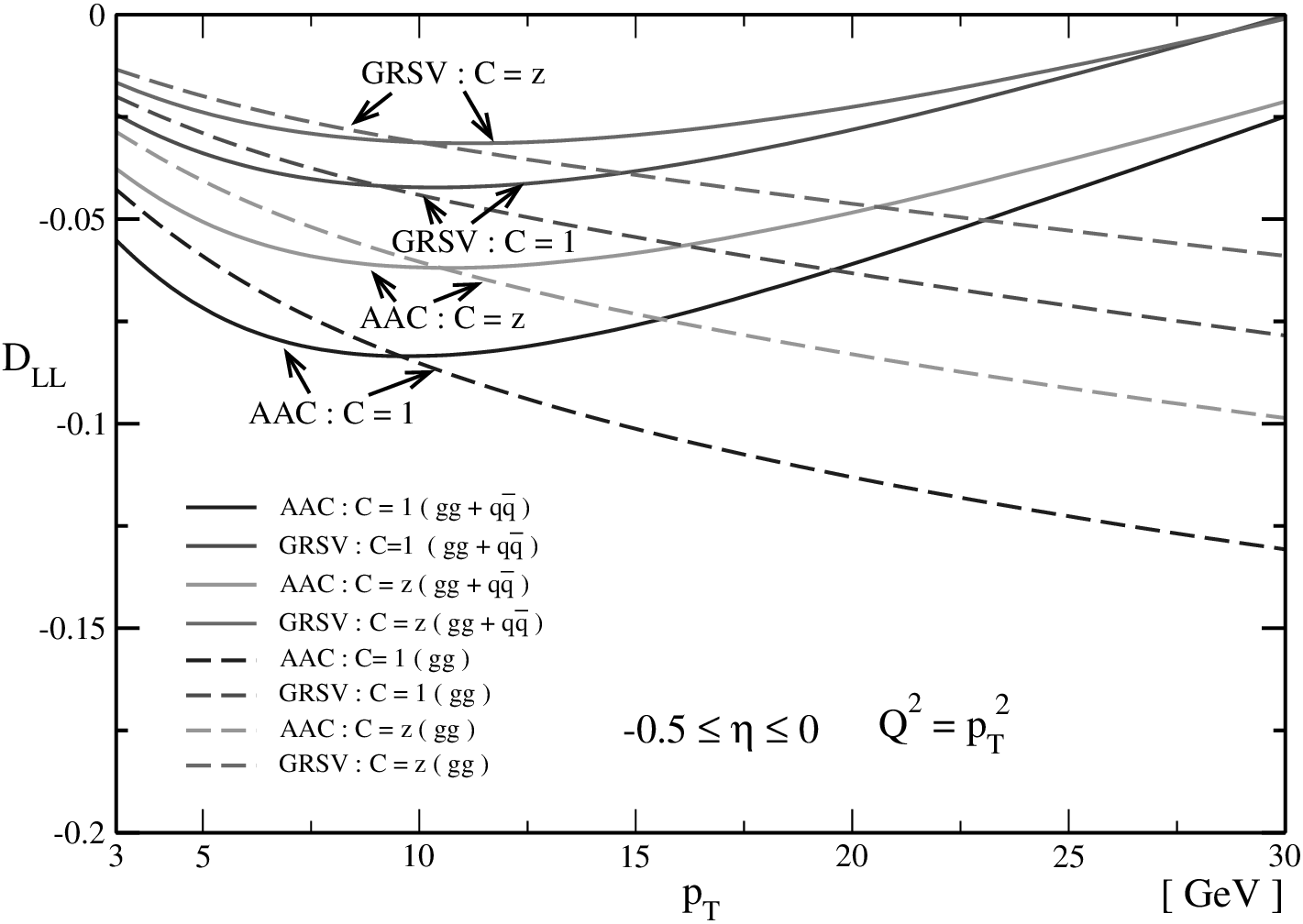}
  \caption{
The same as in Fig.~\ref{d_cross_200_pt},
but for $\Lambda_c^+$ production.
}
\label{l_cross_200_pt}
\end{figure}

\begin{figure}[htb]
\vspace*{0.5cm}
\includegraphics[scale=0.28]{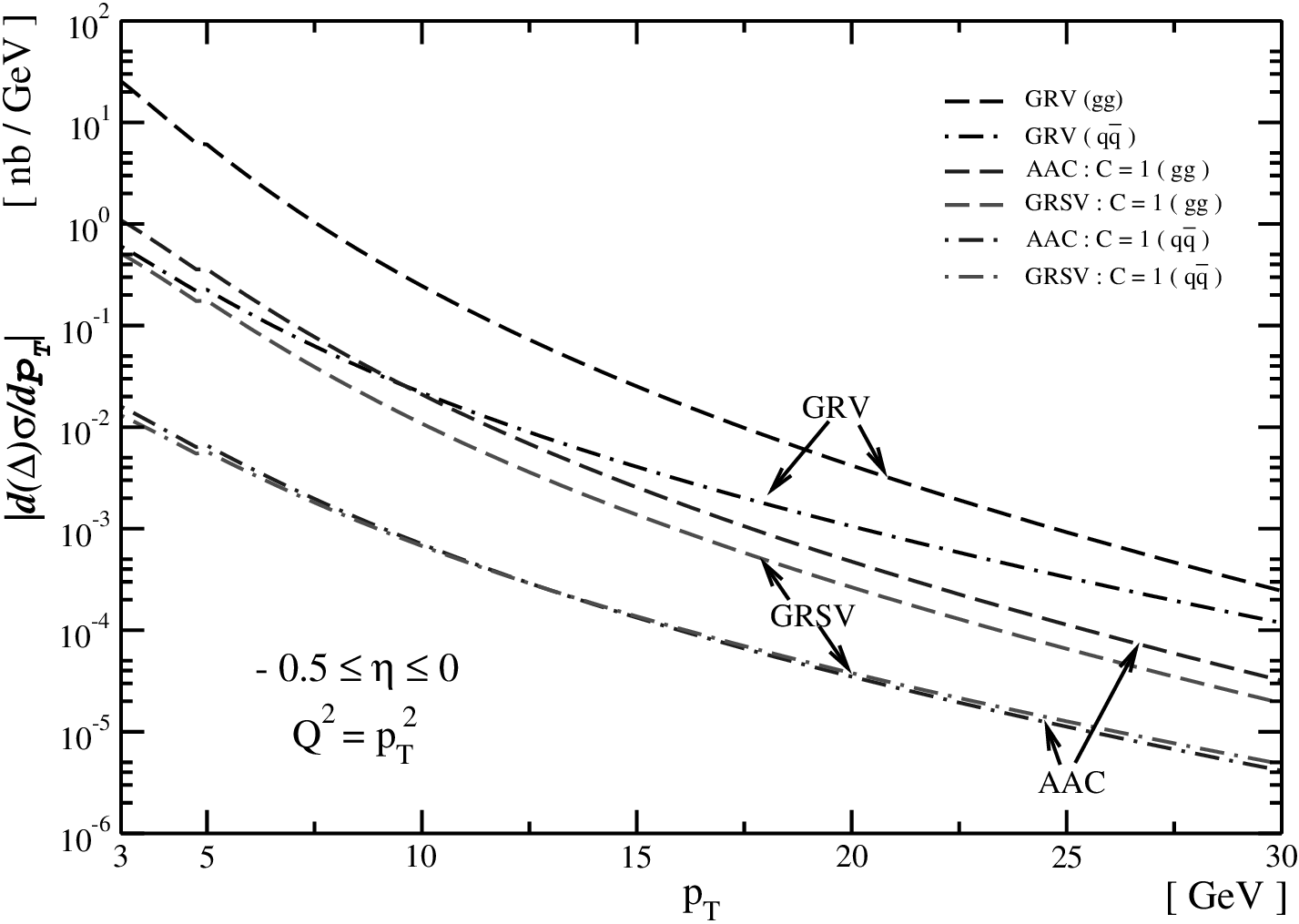}
\hspace*{0.2cm}
\includegraphics[scale=0.28]{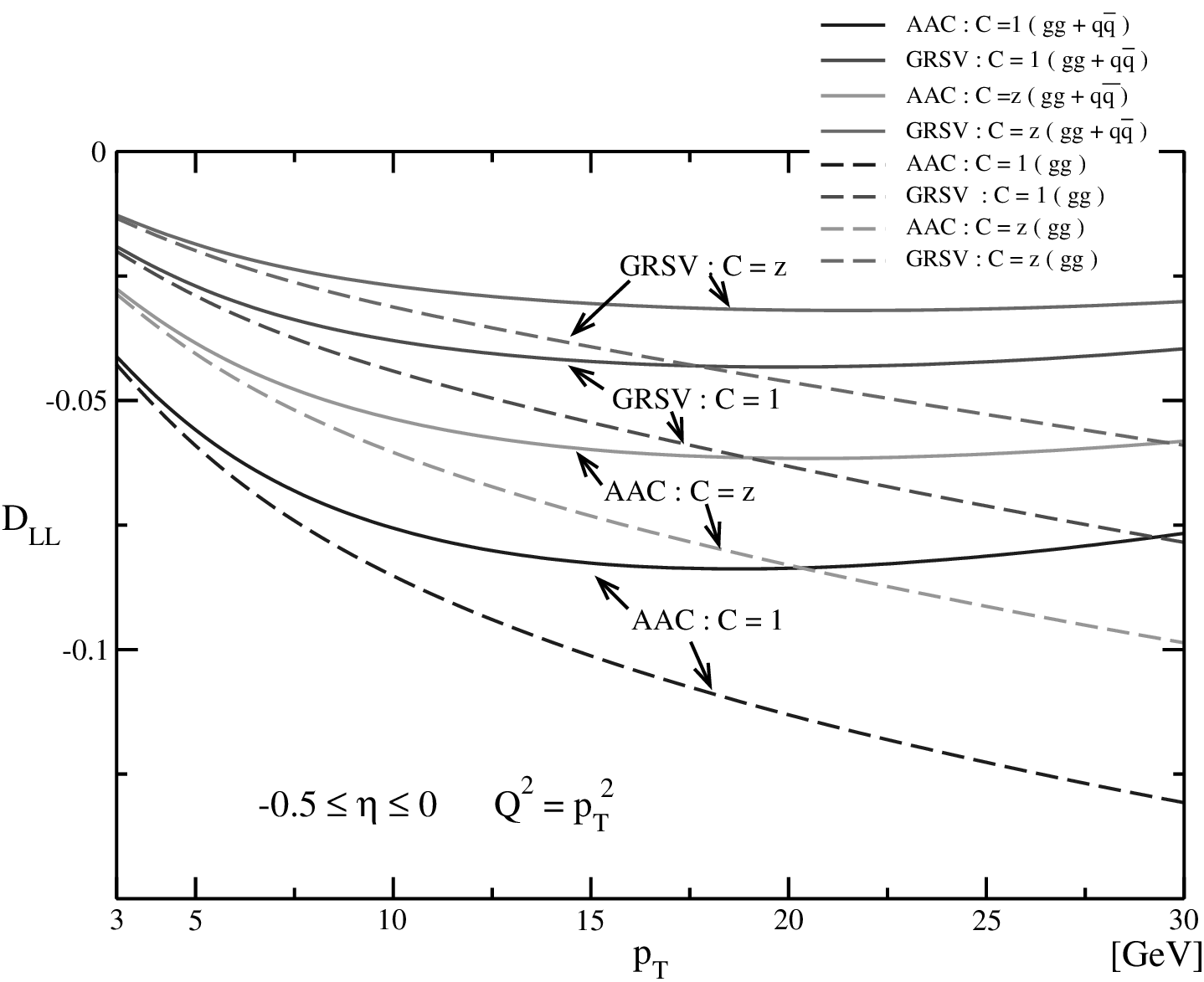}
  \caption{
The same as in Fig.~\ref{d_cross_500_pt},
but for $\Lambda_c^+$ production.
}
\label{l_cross_500_pt}
\end{figure}

Since the charm quark is heavy, the Bjorken--$x$ of the parton
taking part in charm quark pair production 
in the subprocess is not very large 
for $pp$ collisions with an appropriate energy.  
In this case, a charm quark is expected 
to be dominantly produced via gluon--gluon fusion because of the
rather large gluon distribution. 
However, since valence quark densities become larger
in large $x$ region, we could not neglect a contribution 
from a quark--anti-quark annihilation $q\bar{q}\to c\bar{c}$ 
in high $\pt$ region.
Therefore, in order to get a good information on the gluon
polarization, we must find the kinematical region where
$gg \to c\bar{c}$ dominates over $q\bar{q} \to c\bar{c}$.
To do so, we calculated the $\pt$ distribution of the cross
sections $d\Delta\sigma/d{\pt}$, $d\sigma/d{\pt}$
and spin correlation asymmetries $D_{LL}$ 
for $D^{*+}$ and $\Lambda_c^+$ productions
at $\sqrt{s}$=200GeV ($\sqrt{s}$=500GeV) as shown 
in Fig.~\ref{d_cross_200_pt} (Fig.~\ref{d_cross_500_pt}) and
Fig.~\ref{l_cross_200_pt} (Fig.~\ref{l_cross_500_pt}),
respectively.  In the left panel of these figures, we present
the results only for the model (A), since
the purpose of this analysis is to compare a contribution of 
$gg\to c\bar{c}$ with the one of $q\bar{q}\to c\bar{c}$.
Since in the $\pt$ region of $3\leq \pt\leq 5$ GeV, 
the cross section, $d\Delta\sigma/d{\pt}$ and $d\sigma/d{\pt}$, 
of $gg \to c\bar{c}$ is much larger than the one of 
$q\bar{q}\to c\bar{c}$, 
the $D_{LL}$ for the sum of $gg \to c\bar{c}$ 
and $q\bar{q}\to c\bar{c}$ do not change much from the one for
$gg \to c\bar{c}$ alone.  Therefore, we can say 
that a contribution from the
subprocess $q\bar{q}\to c\bar{c}$ can be safely neglected
in the region of $3\leq \pt\leq 5$ GeV.

\begin{figure}[htb]
\vspace*{0.2cm}
\includegraphics[scale=0.28]{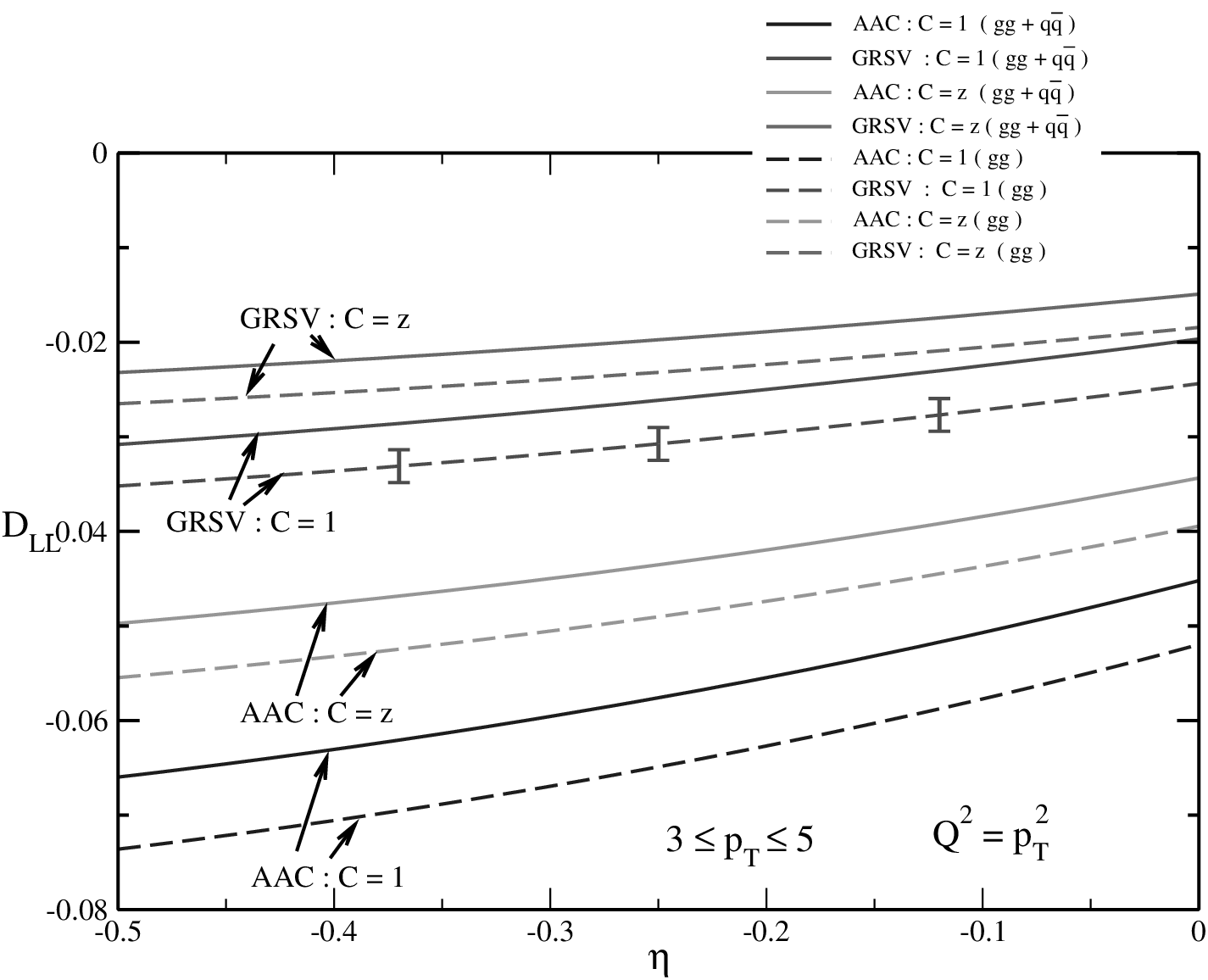}
\hspace*{0.2cm}
\includegraphics[scale=0.28]{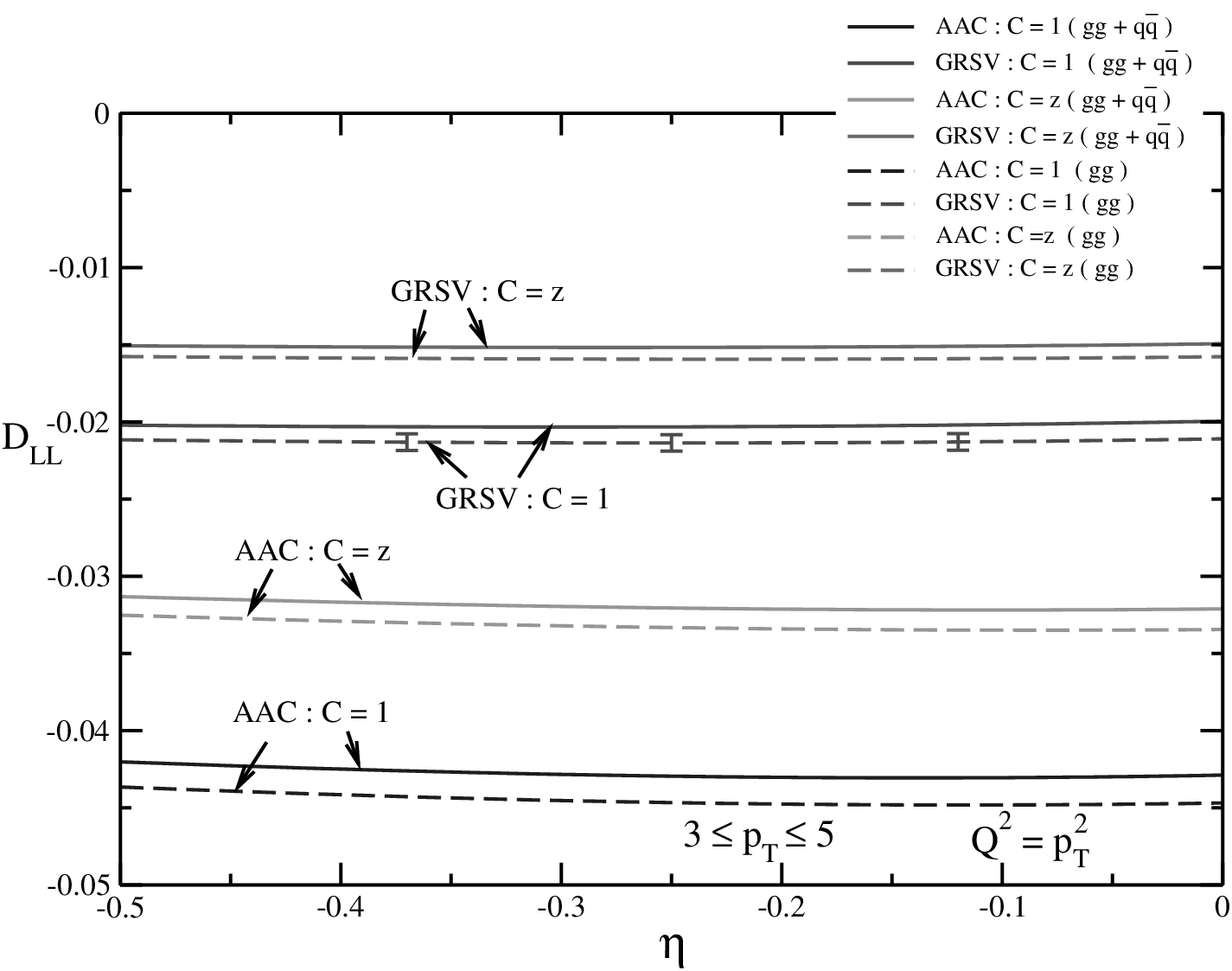}
\caption{The  $\all$ for $\dsp$ productions 
as a function of $\eta$ at 
$\sqrt{s}$=200~GeV(left panel)
and $\sqrt{s}$=500~GeV(right panel)
}
\label{d_dll_eta}
\end{figure}

\begin{figure}[htb]
\vspace*{0.2cm}
\includegraphics[scale=0.28]{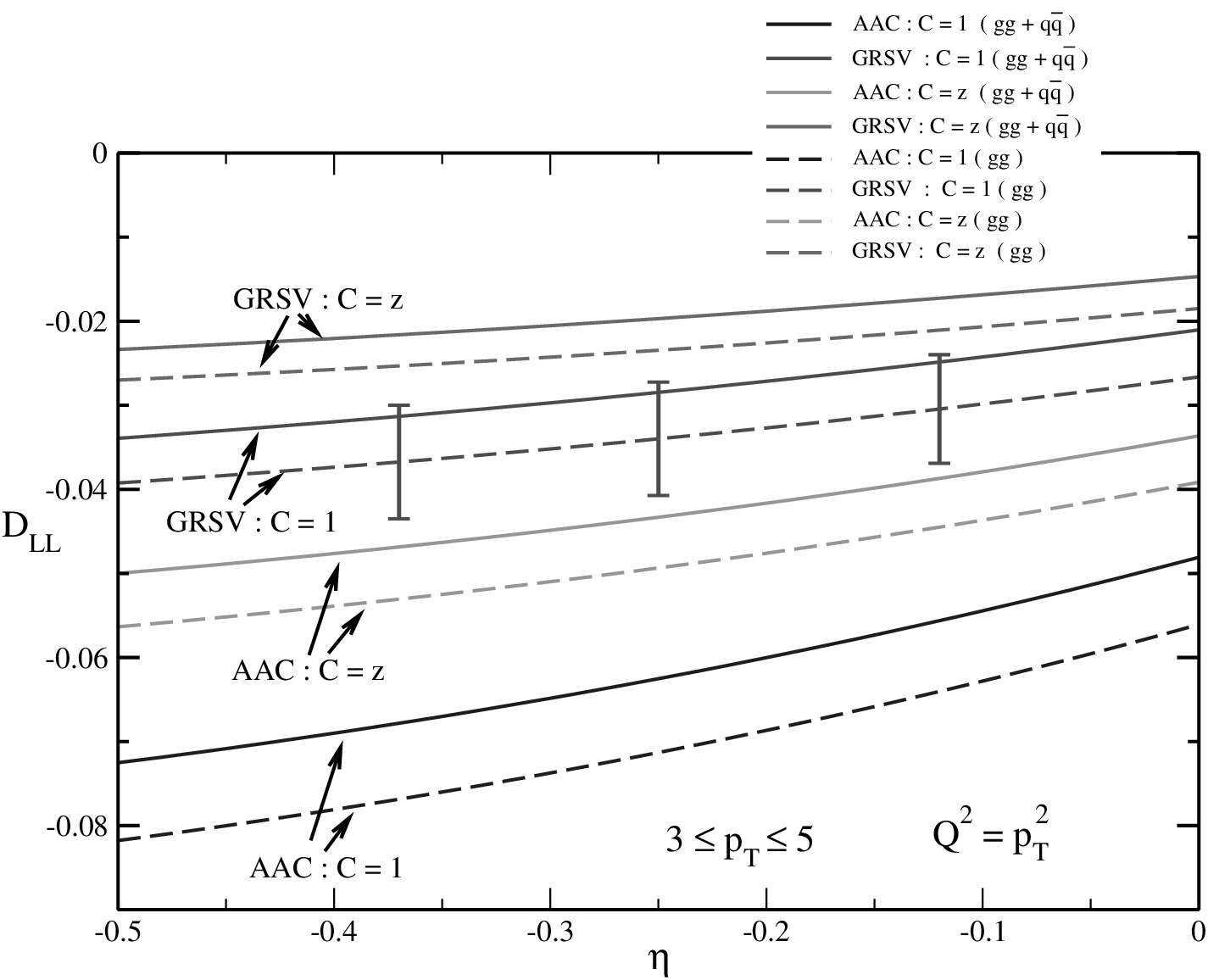}
\hspace*{0.2cm}
\includegraphics[scale=0.28]{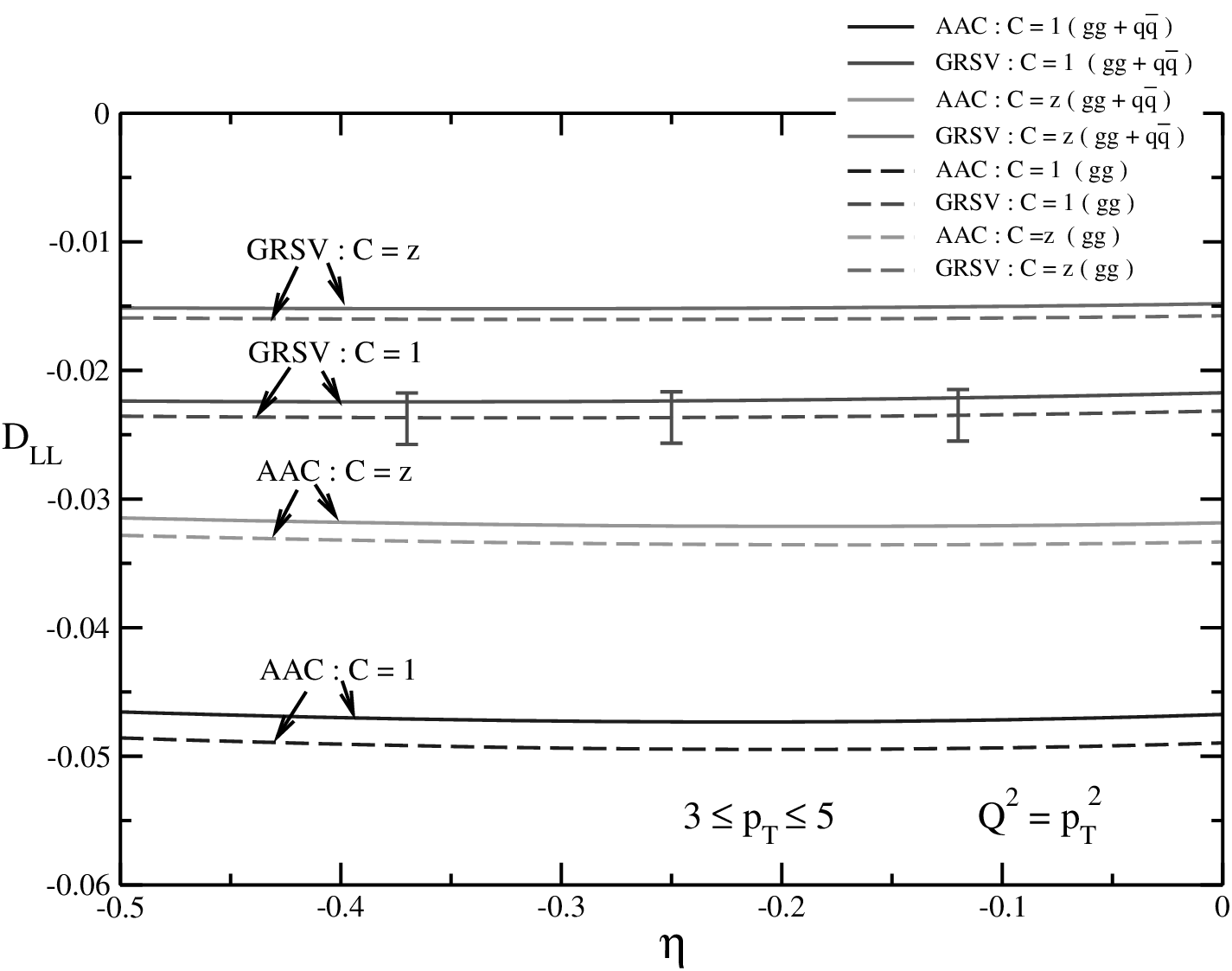}
 \caption{
The $\all$ for $\lmc$ productions as a function of $\eta$ at 
$\sqrt{s}$=200~GeV(left panel)
and $\sqrt{s}$=500~GeV(right panel)
}
\label{l_dll_eta}
\end{figure}

We also calculated $D_{LL}$ as a function of $\eta$ for 3 kinematical
regions of $\eta$, i.e. (i) $-0.5\leq\eta\leq0$, (ii) 
$-0.5\leq\eta\leq0.5$, and (iii) $0\leq\eta\leq0.5$.
However, since the result for the region (i) distinguishes most
clearly a contribution of  $q\bar{q}\to c\bar{c}$
from the one of $gg\to c\bar{c}$, 
we show only the result for the region (i) in 
Fig.~\ref{d_cross_200_pt} $\sim$ Fig.~\ref{l_cross_500_pt}.
For this limited $\pt$ region ($3\leq \pt\leq 5$ GeV), 
$D_{LL}$ for $\dsp$ and $\lmc$ 
production are shown 
in Fig.~\ref{d_dll_eta} $\sim$ Fig.~\ref{l_dll_eta}, respectively,
at $\sqrt{s}=200$GeV and $\sqrt{s}$=500GeV, where we see a rather
big model-dependence of polarized gluons and also the spin-dependent
fragmentation functions.

Furthermore, to know how $D_{LL}$ is sensitive to the behavior
of polarized gluons in the proton, we calculated the statistical
sensitivities of $D_{LL}$, i.e. $\delta \all$, which is defined
by the following formula~\cite{sudoh};
\begin{equation}
\delta \all \simeq \frac{1}{P |\alpha|}
\sqrt{\frac{3}{b_{h_c}~\epsilon~L~\sigma}}.
\label{dall}
\end{equation}
To numerically estimate the value of $\delta \all$, 
we used the following parameters:
the beam polarization; $P=$70\%,
a integrated  luminosity; 
$L$=320pb$^{-1}$(800pb$^{-1}$) for
$\sqrt{s}=200~(500)~$GeV~\cite{rhic},
the trigger efficiency; $\epsilon =10\%$ for detecting produced
charmed hadron events, decay asymmetry parameter; $\alpha=1.0~ (-0.98)$
for $D^{*+}$ ($\Lambda_c^+$) decay
and
a branching ratio;
$b_{\dsp}\equiv {\rm Br}(\dsp \to D^0 \pi^+)
 {\rm Br}(D^0 \to K^-\pi^+)
\simeq 2.5\times 10^{-2} ,~
b_{\Lambda_c^+}\equiv {\rm Br}(\lmc \to \Lambda \pi^+)
 {\rm Br}(\Lambda \to p\pi^-)\simeq 5.8\times 10^{-3}
$~\cite{pdg}~\footnote{
In order to determine the polarization of $D^{*+}$ and 
$\Lambda_c^+$, observation of the angular distribution 
for decay channels of these particles is necessary.
Practically this could be done 
by detecting charged particles
in final the state. 
}.
$\sigma$ denotes the unpolarized cross section integrated over a
corresponding $\eta$ region. 
The factor $3$ in Eq. (\ref{dall}) is an acceptance factor
for our processes~\cite{sudoh}.
In Figs.~\ref{d_dll_eta} and \ref{l_dll_eta},
statistical sensitivities, $\delta \all$, are attached only
to the dashed line of $\all$ which 
were calculated using the GRSV01 parametrization model of 
polarized partons and the non-relativistic fragmentation model (model (A))
\footnote{Note that as shown in Eq.(\ref{dall}), $\delta D_{LL}$ 
does not depend on both the model of polarized gluons and the model 
of fragmentation functions.}.  As shown here, statistical sensitivities 
$\delta \all$ are so small that the processes must be feasible
for measuring the $D_{LL}$.
 
From these results, we see that the $\eta$ distributions of $\all$ 
are effective for testing the model of not only polarized 
gluon distributions but also spin-dependent fragmentation functions 
for all cases ($\dsp$ production and $\lmc$ production) calculated 
at $\sqrt{s}=$200 GeV and 500 GeV. 
Especially, $\eta$ distributions of $\all$ for $\dsp$ productions are quite
promising, though the magnitude of $\all$ becomes
smaller with increasing center of mass energy from 
$\sqrt{s}=$200 GeV to 500 GeV. 
 
%
%
%
%
%
%
%
%
%
%
%
\section{Conclusion}

In order to get information about polarized gluons in the
proton, we proposed two charmed hadron production processes,
i.e. polarized $\dsp$ meson productions and polarized $\lmc$ 
baryon productions, 
which will be observed in the forthcoming RHIC experiments. 
As described in Introduction, the processes contain two
production mechanisms, gluon--gluon fusion ($gg \to c\bar{c}$) and
quark--anti-quark annihilation ($q\bar{q} \to c\bar{c}$).  
Thus, to study the gluon polarization in the proton, 
it is necessary to find the kinematical region where
$gg \to c\bar{c}$ dominates over $q\bar{q} \to c\bar{c}$.

From the numerical calculation at the lowest order of QCD,
we found that the $\eta$ distribution of $\all$ 
in the limited $\pt$ region ($3\leq \pt \leq 5$ GeV) is
quite promising for testing not only the 
model of polarized gluons in the proton but also the model of 
spin-dependent fragmentation functions.
However, it should be noted that the assumption on 
the spin-dependent fragmentation function might be 
somewhat too simple in the present analysis.  
In order to obtain more reliable prediction, further 
investigation on spin-dependent fragmentation functions is necessary.
For the $\Lambda$ baryon production,
similar analysis was performed by
D. de Florian {\it et. al}~\cite{delall}.
Our analysis might be complementary to their analysis but 
more effective for extracting information
of the polarized gluon in the proton, since the non-relativistic 
quark model works better for $\lmc$ ($\dsp$)  than $\Lambda$ 
and furthermore the separation of 
gluon fusion and $q\bar{q}$ annihilation is easier for our processes. 

Though the present calculation is confined to the leading order,
the results are interesting and we hope that our analysis could be 
tested in the forthcoming RHIC experiments.

\section*{Acknowledgment}
We would like to thank T. Iwata and N. Saito 
for useful discussions and comments on experimental feasibility.
We are very much thankful to K. Sudoh and S. Oyama for interesting
discussions and advice at early stage of this work.
We thank K. Sudoh for informing us a newly improved expression 
of eq.(12) for estimating statistical uncertainties.
One (K.O.) of authors is grateful to T. Yamanishi, K. Sasaki and T. Ueda
for giving him useful information related to this work.
K. O. is  supported by a Grant-in-Aid for Young Scientists (B)
from the Ministry of Education, Culture, Sports, Science and Technology 
of Japan(\#17740157).

\end{document}